\documentclass[useAMS,usenatbib]{mn2e}
\usepackage{graphicx, amsmath, amssymb,epsfig}

\def\kms{\mbox{km~s$^{-1}$}}
\def\kpc{\mbox{kpc}}

\def\kpch{\mbox{$h^{-1}$kpc}}

\def\LCDM{\mbox{$\Lambda$CDM}}

\def\Mpch{\mbox{$h^{-1}$Mpc}}
\def\Mvir{\mbox{$M_{\rm vir}$}}
\def\M200{\mbox{$M_{\rm 200}$}}

\def\Msunh{\mbox{$h^{-1}M_\odot$}}

\def\Rvir{\mbox{$R_{\rm vir}$}}
\def\R200{\mbox{$R_{\rm 200}$}}

\def\Vmax{\mbox{$V_{\rm max}$}}
\def\V200{\mbox{$V_{\rm 200}$}}

\def\Vrad{\mbox{$\langle V_r\rangle$}}

\def\Omatter{\mbox{$\Omega_{\rm matter}$}}

\newcommand{\apjs}{ApJ Suppl.}
\newcommand{\apj}{ApJ}

\newcommand{\mnras}{MNRAS}

\title[$\LCDM$ halo concentrations]{Halo concentrations in the standard $\LCDM$ cosmology}
\author[F. Prada, et al.]{Francisco~Prada$^{1,2}$\thanks{E-mail: fprada@iaa.es}, 
Anatoly~A.~Klypin$^{3}$, Antonio~J.~Cuesta$^{1,4}$,
\newauthor
Juan~E.~Betancort-Rijo$^{5}$ and Joel~Primack$^6$\\
 \vspace{-0.2cm}\\
$^{1}$Instituto de Astrofisica de Andalucia (CSIC), E-18080 Granada, Spain\\
$^{2}$Visiting Professor at the  Institute for Computational Cosmology, Department of Physics, University of Durham, UK\\
$^{3}$Astronomy Department, New Mexico State University, Las Cruces, NM, USA\\
$^{4}$Present address: Yale Center for Astronomy and Astrophysics, Yale University, New Haven, CT, USA\\
$^{5}$Instituto de Astrofisica de Canarias, Tenerife, Spain \\
$^{6}$Department of Physics, University of California at Santa Cruz, Santa Cruz, CA, USA}
\begin{document}

\date{Submitted, 2011 April 27}

\pagerange{\pageref{firstpage}--\pageref{lastpage}} \pubyear{2002}

\maketitle

\label{firstpage}

\begin{abstract}

  We study the concentration of dark matter halos and its evolution in
  N-body simulations of the standard $\LCDM$ cosmology. The results
  presented in this paper are based on 4 large $N$-body simulations
  with $\sim$ 10 billion particles each: the 
  Millennium-I and II, Bolshoi, and MultiDark 
  simulations. The MultiDark (or BigBolshoi) simulation is introduced in this paper.  
  This suite of simulations with
  high mass resolution over a large volume allows 
  us to compute with
  unprecedented accuracy the concentration over a large range of
  scales (about six orders of magnitude in mass), which constitutes
  the state-of-the-art of our current knowledge on this basic property
  of dark matter halos in the $\LCDM$ cosmology. 
  We find that there is consistency
  among the different simulation data sets, despite the
  different codes, numerical algorithms, and halo/subhalo finders used
  in our analysis.
  We confirm a novel
  feature for halo concentrations at high redshifts: a flattening and
  upturn with increasing mass. The concentration $c(M,z)$ as a
  function of mass and the redshift and for different cosmological
  parameters shows a remarkably complex pattern. However, when
  expressed in terms of the linear $rms$ fluctuation of the density
  field $\sigma(M,z)$, the halo concentration $c(\sigma)$ shows a
  nearly-universal simple U-shaped behaviour with a minimum at a well
  defined scale at $\sigma\sim 0.71$. Yet, some small
  dependences with redshift and cosmology still remain. At the
  high-mass end ($\sigma < 1$) the median halo
  kinematic profiles show large signatures of infall and highly radial
  orbits. This $c$--$\sigma(M,z)$ relation can be accurately
  parametrized and provides an analytical model for the dependence of
  concentration on halo mass. When
  applied to galaxy clusters, our estimates of concentrations are
  substantially larger -- by a factor up to 1.5 -- than previous results from
  smaller simulations,
  and are in much better agreement with results of observations.

\end{abstract}


\section{Introduction}

N-body cosmological simulations have been essential for understanding
the growth of structure in the Universe, and in particular, they have
been crucial for studying the properties of dark matter halos in the
standard Lambda Cold Dark Matter ($\LCDM$) cosmology.  Simulations are
also an invaluable tool for analyzing galaxy surveys,  
for studying the
abundance evolution of clusters of galaxies, and 
for semi-analytical models of galaxy formation. In recent years, the
development of numerical codes and access to powerful supercomputers
have made it
possible to perform $\it Grand\, Challenge$ cosmological
simulations with high mass resolution over a large volume, which
provide the basis to attack many problems in cosmology. The
Millennium simulation \citep[][MS-I]{Springel05}, its smaller volume and 
higher-resolution
Millennium-II version \citep[][MS-II]{BK09} and the new Bolshoi simulation
\citep[][]{Klypin10}, in this respect, constitute a remarkable
achievement.

Millennium, Bolshoi, and the new MultiDark (or BigBolshoi) simulation,
which is introduced in this work, allow us to estimate for the $\LCDM$
cosmology, with unprecedented statistics, the concentration of dark
matter halos and its evolution over six orders of magnitude in
mass. The comparison between this suite of cosmological simulations
also provides an unique opportunity to study consistency on dark
matter halo statistics using different codes, numerical algorithms,
halo finders and cosmological parameters. Halo concentrations
$c\equiv\Rvir/r_s$ have been studied extensively during the last
decade. 
Here $\Rvir$ is the virial radius of a halo and $r_s$ is the break radius
between an inner $\sim r^{-1}$ density profile and an outer $r^{-3}$ profile.
Typically, the median halo concentration declines with
increasing mass and redshift \citep{Bullock01}, 
and the shape of the mass--concentration
median relation evolves. Recently, \citet[][]{Klypin10} found a novel
feature: at high redshift, the concentration flattens, and then
increases slightly for high masses. Numerous works made use of
cosmological simulations to determine the correlation and scatter of
concentration with halo mass, its evolution, and its dependence on environment
and cosmology \citep[e.g.][]{NFW,Bullock01, Eke01, Wechsler02, Zhao03a, Neto07,
  Gao08, Maccio08, Zhao09, MC10, Klypin10}. Moreover, the dependence
of halo concentrations on their merging history has been studied also
in detail. For example, \citet[][]{Wechsler02,Zhao03a,Zhao09}
determine concentrations using accurate modeling of halo mass
accretion histories.  Simulations and analytical models have tried
also to understand the formation and evolution of the central density
cusp of dark matter halos - assuming that they grow inside out during
the different phases in their mass accretion histories - to provide
predictions on halo concentrations \citep[see
e.g.][]{Reed05,Lu06,Romano07,Yehuda07,Salvador07}. The origin of halo
concentrations ultimately is linked to the physical process that yield
the formation of the NFW density profiles in the cold dark matter
scenario \citep[see also][]{Dalal10}.

Here, we show that the dependence of concentration on halo mass and
its evolution can be obtained from the $rms$ 
fluctuation 
amplitude of the linear density
field $\sigma(M,z)$ by using our new analytical model based on the
concentration--$\sigma(M,z)$ relation. This model is able to reproduce
all relevant features observed in the halo mass--concentration median
relations at different redshifts: the decline of concentration with
mass, its flattening and upturn.

In this paper, we study the evolution of halo concentrations in N-body
cosmological simulations of a $\LCDM$ cosmology. We cover a large
range of scales, going from halos hosting dwarf galaxies to massive
galaxy clusters.  This corresponds to halo maximum circular velocities
ranging from 25 to 1800 $\kms$, covering six orders of magnitude in
mass. In Section 2 we describe the simulations and halo catalogs used
in this work.  Methodology to estimate halo concentrations is
presented in section 3. 
In this paper we analyze only distinct halos. A halo is called
distinct if its center is not inside the virial radius of a larger halo. 
In Section 4 we study concentrations of distinct halos 
and their evolution with redshift using data from the
Millennium, Bolshoi and the new MultiDark simulations. A new
parameterization of the distinct halo concentrations as a function of
the $rms$ of the linear density field $\sigma(M,z)$ is presented in
Section 5.  Main results of this work are summarized in Section
5. An Appendix gives discussion of various selection effects.

\section{Simulations and halo identification}

Table 1 summarizes the basic numerical and cosmological parameters of
the four simulations used in this work. The Millennium simulation
\citep[][]{Springel05} and the more recent Millennium-II
simulation \citep[][]{BK09} adopted the same WMAP first
year cosmological parameters, used the same number of particles to
resolve the density field, and share the same output data
structure. MS-II was done in a cubic simulation box one-fifth the linear size
of the original MS-I with 5 times better force resolution and 125 times
better mass resolution. MS-I,II were run with the TREE-PM
codes GADGET-2 and GADGET-3 respectively.

The Bolshoi \citep[][]{Klypin10} and the MultiDark
simulations use the latest WMAP5 and WMAP7 cosmological parameters,
which are also consistent with other recent observational
constraints
-- see Figure 1 of \citet[][]{Klypin10} . 
The Adaptive-Refinement-Tree (ART) code was used for these
two simulations. ART is an Adaptive-Mesh-Refinement (AMR) type code. A
detailed description of the code is given in \citet{Kravtsov97} and
\citet{Kravtsov99}.  The code was parallelized using MPI libraries and
OpenMP directives \citep{Gottlober08}. Details of the time-stepping
algorithm and comparison with GADGET and PKDGRAV codes are given in
\citet{Klypin09}.  The ART code increases the force resolution by
splitting individual cubic cells into $2\times 2\times 2$ cells with
each new cell having one half the size of its parent.  This is done
for every cell if the density of the cell exceeds some specified
threshold. The value of the threshold varies with the level of
refinement and with the redshift and is typically 2-5 particles per
cell. 180 snapshots were saved for analysis. Details of the Bolshoi
simulations are given in \citet{Klypin10}.

 Here we give some details of the MultiDark
run. Initial conditions were set at the redshift $z_{\rm init} =65$
using the same power spectrum as for Bolshoi. The force resolution in
the ART code varies with time. The {\it comoving} resolution (size of the
smallest cell) for the MultiDark was equal to $244\,\kpch$ at $z>10$.
When the fluctuations started to collapse, the resolution became
smaller and at $z=0$ it was $7.6\,\kpch$. The ART code is designed in
such a way that the {\it proper} (physical) resolution is nearly
preserved over time. For MultiDark the proper resolution was $\sim
7\,\kpch$ for $z=0-8$ as compared with $~1\,\kpch$~ for $z=0-20$ for
Bolshoi. The total volume resolved at the highest resolution at $z=0$ was
$360(\Mpch)^3$ and $2.8\times 10^4(\Mpch)^3$ for twice worse
resolution. 50~snapshots were saved for analysis.

 The main difference in cosmological parameters between the simulations is
that the Millennium simulations adopted a substantially larger
amplitude of perturbations $\sigma_8$ that is nearly $4\sigma$ away from 
 recent constraints \citep[e.g.,][]{Klypin10,Guo11}. The
difference is even larger on galaxy scales since the Millennium
simulations used a larger tilt $n_s$ of the power spectra (see Table 
1, and Figure 2 in \citet{Klypin10} for the comparison of the linear
power spectra of Bolshoi and Millennium simulations).

\begin{table*}
\centering
 \begin{minipage}{170mm} 
   \caption{Basic parameters of the cosmological
     simulations. $L_{box}$ is the side length of the simulation box,
     $N_p$ is the number of simulation particles, $\epsilon$ is the
     force resolution in comoving coordinates, $M_p$ refers to the
     mass of each simulation particle, and the parameters $\Omega_m$,
     $\Omega_{\Lambda}$ , $\Omega_b$, $n_s$ (the spectral index of the
     primordial power spectrum), $h$ (the Hubble constant at present
     in units of 100 $km/s Mpc^{-1}$) and $\sigma_8$ (the rms
     amplitude of linear mass fluctuations in spheres of 8 $h^{-1}Mpc$
     comoving radius at redshift $z=0$) are the $\Lambda$CDM
     cosmological parameters assumed in each simulation.}
  \begin{tabular}{@{}lccccclclccr@{}}
  \hline
   Name          &$L_{box}$& $N_p$& $\epsilon$& $M_p$ &$\Omega_m$ &$\Omega_{\Lambda}$ &$\Omega_b$& $n_s$ & $h$ & $\sigma_8$ &reference\\
                 & ($\Mpch$) &               &  ($\kpch$) &  ($h^{-1}M_{\odot}$) &  &  &  &  &  & &  \\
 \hline
 Millennium      & 500  & $2160^3$ & 5  & $8.61\, 10^8$ & 0.25 & 0.75 & 0.0450 & 1.00 & 0.73  & 0.90 & Springel et al. 2005   \\
 Millennium-II   & 100  & $2160^3$ &  1 & $6.89\, 10^6$ & 0.25 & 0.75 & 0.0450 & 1.00 & 0.73  & 0.90 & Boylan-Kolchin \\
                 &      &          &    &               &      &      &        &      &       &      & \,\, et al. 2009 \\
 Bolshoi         & 250  & $2048^3$ &  1 & $1.35\, 10^8$ & 0.27 & 0.73 & 0.0469 & 0.95 & 0.70  & 0.82 & Klypin et al. 2010 \\
 MultiDark       & 1000 & $2048^3$  &   7  & $8.63\, 10^9$  &  0.27 & 0.73 & 0.0469 & 0.95 & 0.70  & 0.82  &    this work \\
\hline
\end{tabular}
\end{minipage}
\end{table*}

In this paper we analyze only distinct halos. A halo is called
distinct if its center is not inside the virial radius of a larger
halo. One of the most important characteristics of a halo is its
maximum circular velocity:
  \begin{equation}
    V^2_{\rm max}=\max\left[\frac{GM(<r)}{r}\right].
\end{equation}
There are several advantages of using $\Vmax$ 
to characterize a halo
as opposed to the virial
mass. First, $\Vmax$ does not have the ambiguity related with the
definition of ``virial mass.'' Virial mass and radius vary depending
on overdensity threshold used. For the often-employed overdensity 200
and ``virial'' overdensity thresholds, the differences in
definitions result in changes in the halo radius from one definition
to another and, thus, in concentration, by a factor of 1.2-1.3, with the
exact value being dependent on the halo concentration. Second and more
important, the maximum circular velocity $\Vmax$ is a better quantity to
characterize halos when we relate them to the galaxies inside these halos. 
For galaxy-size halos the maximum circular
velocity is defined at a radius of $\sim 40$~kpc: closer
to sizes of luminous parts of galaxies than the much larger virial
radius, which for the Milky-Way halo is $\sim 250~\kpc$
\citep[e.g.,][]{Klypin02}.

In both Millennium simulations dark matter halos were found using the
friends-of-friends algorithm \citep[FOF,][]{Davis85}, with a linking
length of $b=0.2$. The SUBFIND code \citep{Springel01} searched for
bound substructures within every FOF group. In most cases, the main
component of a FOF group is typically a dominant spherical ``subhalo''
whose particles make up most of the bound component of the FOF
group. This dominant or central subhalo can be identified as a
distinct halo with a maximum circular velocity $\Vmax$.
For our analysis we consider all distinct halos (i.e. dominant
subhalos of FOF groups) with maximum circular velocities greater than
$\Vmax>25\,\kms$ and $\Vmax>70\,\kms$ for MS-II and MS-I
respectively. Halo velocity functions are complete above these
velocities \citep[see][]{Springel05,BK09}. We have
downloaded\footnote{FOF and SUBFIND halo catalogs from MS-I and MS-II
  are available at http://www.g-vo.org/MyMillennium3 .} MS-I and MS-II
halo catalogs for several epochs up to $z=10$, as close as possible to
those redshifts available for Bolshoi and MultiDark.

\begin{figure}
\centering
\includegraphics[width=0.45\textwidth]{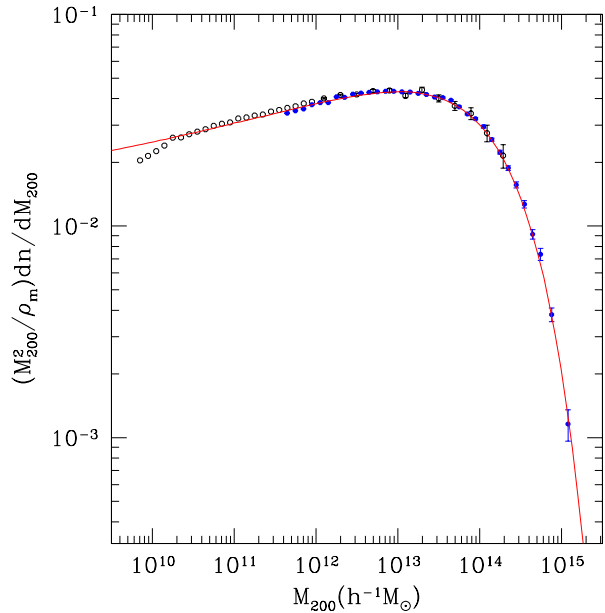}
\caption{Mass functions of distinct halos in Bolshoi (open circles)
  and MultiDark (filled circles) simulations at $z=0$. The analytical
  approximation of \citet{Tinker08} is shown by the curve. In the
  overlapping region  both simulations produce nearly
  identical results.}
\label{fig:Mass}
\end{figure}

Dark matter halos are identified in Bolshoi and MultiDark with a
parallel version of the Bound-Density-Maxima (BDM) algorithm
\citep{KH97}. The BDM is a Spherical Overdensity (SO) code. It
finds all density maxima in the distribution of particles using a
top-hat filter with 20 particles. For each maximum the code estimates
the radius within which the overdensity has a specified value. Among
all overlapping density maxima the code finds one that has the deepest
gravitational potential. The position of this maximum is the center of a distinct halo\footnote{Halo catalogs for these simulations are available at http://www.multidark.org.}. 

The halo radius can be defined as the radius of a sphere within which the
average density is $\Delta$ times larger than the critical density of
the Universe:
\begin{equation}
  M =\frac{4\pi}{3}\Delta\rho_{\rm cr}(z)R^3, \label{eq:Delta} 
\end{equation}
where $\rho_{\rm cr}(z)$ is the critical density of the Universe.  In
this paper we  use the overdensity $\Delta=200$
threshold. Corresponding values for mass and radius are $M=\M200$ and
$R=\R200$. 

\begin{figure}
\centering
\includegraphics[width=0.46\textwidth]{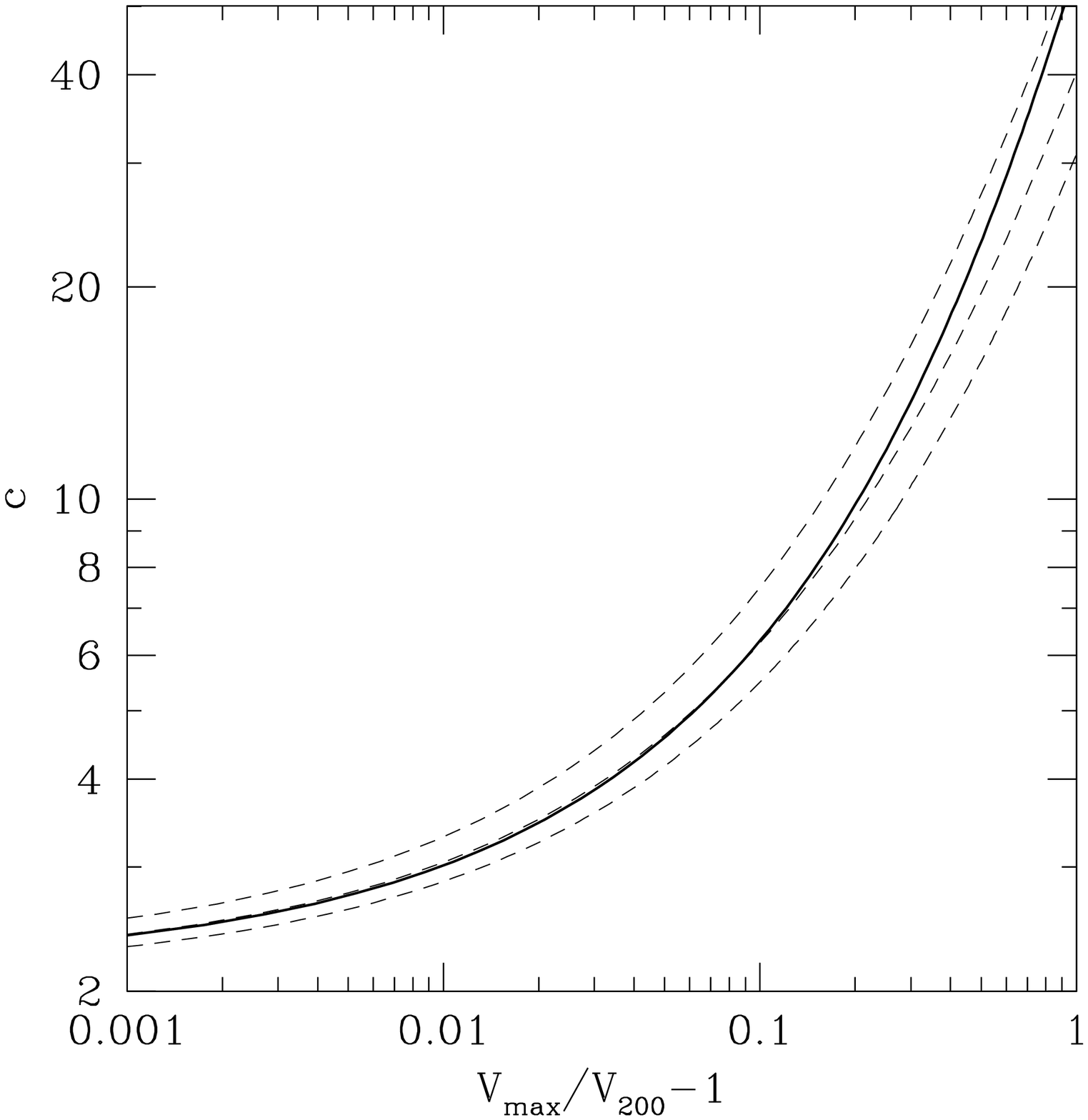}
\caption{Relation between halo concentration $c$ and the ratio of the
  maximum circular velocity $\Vmax$ to the circular velocity at the
  virial radius $\V200$ for the NFW profile (full curve) and for three
  Einasto profiles with parameters $\alpha= 0.15, 0.20, 0.25$ (from top
  to bottom).}
\label{fig:VratC}
\end{figure}

\begin{figure*}
\centering
\includegraphics[width=0.8\textwidth]{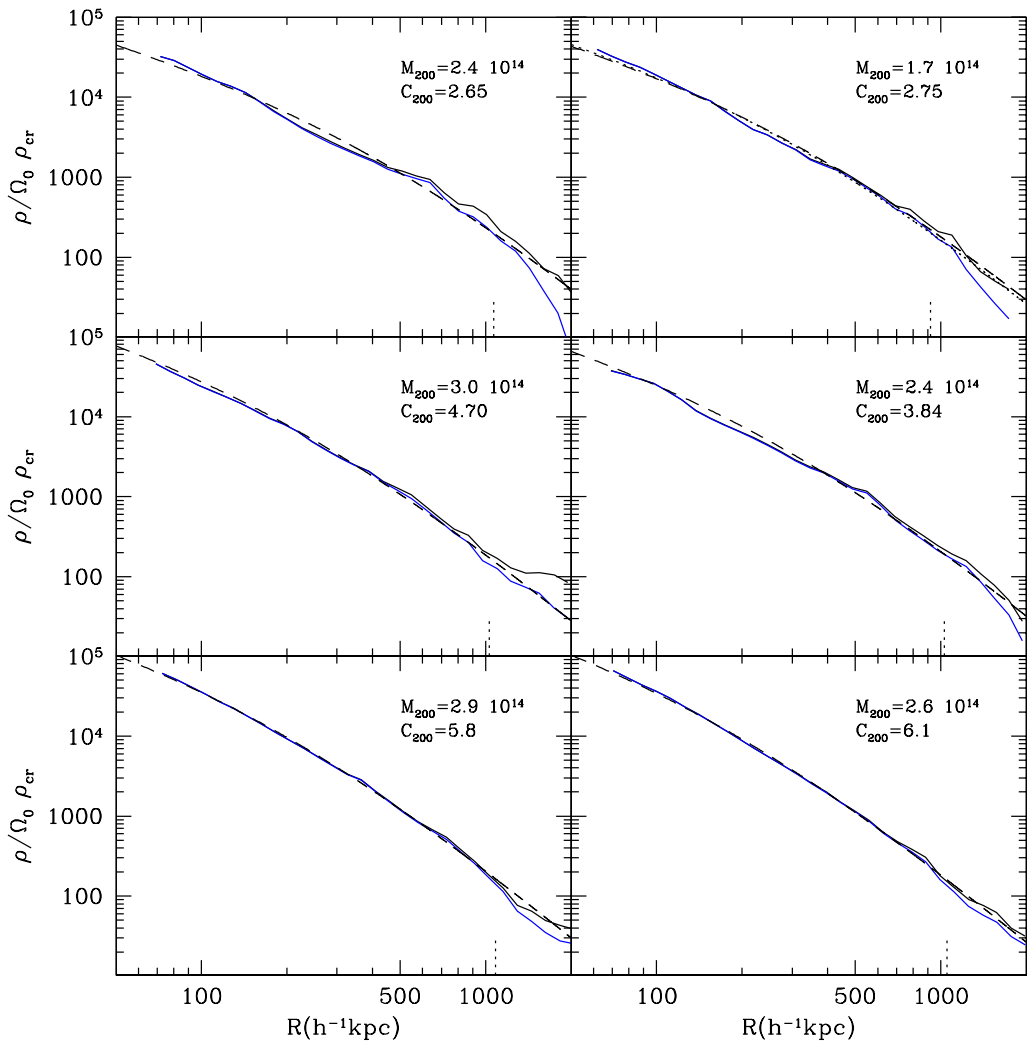}
\caption{Examples of halo density profiles (full curves) and their fits (dashed curves) using
  concentrations obtained with the ratios of the maximum circular
  velocity $V_{\rm max}$ to the halo velocity $V_{200}$. Each panel
  shows two full curves: the density profile of all particles (upper
  curve) and only bound particles (lower curve). Vertical dotted lines show the outer radius of bound
  particles.  Typically the highly concentrated and relaxed halos are
  well fit by the approximations (two bottom panels). Low concentrated
  halos and halos with large perturbations (two top panels) show large
  deviations between the data and approximations. Fitting those
  unrelaxed profiles with standard minimization technique does not
  improve the accuracy of the fits as indicated by dotted curve in
  the top-right panel.}
\label{fig:Profiles}
\end{figure*}

Figure~\ref{fig:Mass} shows differential mass functions of distinct
halos in Bolshoi and MultiDark simulations. The two simulations
demonstrate remarkably good consistency in the overlapping region of
masses $(10^{12}-10^{14})\Msunh$. To describe the mass function
$dn/dM$ we use the following equations:

\begin{eqnarray}
 \frac{dn}{dM} &=& f(\sigma) \frac{\rho_m}{M}\frac{d\log\,\sigma^{-1}}{dM}, \label{eq:dndm}\\
  f(\sigma) &=& A\left[1+\left(\frac{\sigma}{b}\right)^{-a}
                 \right]\exp\left(-\frac{c}{\sigma^2}\right), \label{eq:fsigma}\\
                \phantom{mm} \sigma^2(M,z) &=& \frac{1}{2\pi^2}\int_0^\infty P(k,z)W^2(k,M)k^2dk,
\label{eq:sigma}
\end{eqnarray}
where $\rho_m$ is the cosmological matter density, and $P(k,z)$ is the power
spectrum of fluctuations. 
 Here $\sigma(M,z)$ is the linear rms
fluctuation of density field on the scale $M$ as estimated with the
top-hat filter, whose Fourier spectrum is $W(k,M)$. Parameters of the
approximation are nearly the same as in \citet{Tinker08} for the same
overdensity $\Delta$: $A=0.213$, $a=1.80$, $b=1.85$, and $c=1.57$.

Analysis of halo concentrations is done for different populations of
halos. We limit our analysis to halos containing more than 500
particles. Most of the time we analyze all such halos regardless of
their degree of virialization or interaction.  Occasionally we select
only ``relaxed'' halos. A number of parameters is involved in this
selection.  We define an offset parameter $x_{\rm off}$ as the
distance between the center of a halo (density maximum with the deepest
gravitational potential) and the center of mass of the halo in 
units of the 
virial radius. For each halo we also use the spin parameter
$\lambda$ and the virial ratio $2K/|U|-1$, where $K$ and $U$ are the
total kinetic and potential energies. A halo is called ``relaxed'' if
it satisfies the following three conditions: $x_{\rm off}< 0.1$,
$2K/|U|-1<0.5$, and $\lambda<0.1$. \citet{Neto07} and \citet{Maccio08}
give detailed discussion of these conditions.

\section{Finding halo concentrations}

There are different ways to estimate halo concentration. Traditionally
it is done 
first 
by fitting a NFW profile
\begin{equation}
\rho_{\rm NFW} = \frac{4\rho_s}{x(1+x)^2}, \quad x\equiv \frac{r}{r_s}
\end{equation}
to the spherically averaged density profile of a halo, where $\rho_s$ and $r_s$ are
the characteristic density and radius. 
Then the halo concentration is found as the ratio of the virial radius
to the characteristic radius $r_s$:
\begin{equation}
     c = \frac{\R200}{r_s}
\end{equation} 
 However, the fitting can be difficult. For example, special
attention should be paid to the central region: if fitting starts too
close to the center where the resolution is not sufficient, the
fitting will produce too low concentration 
because the density in the center is underestimated.
 In addition to the
problems with the center and the necessity to deal with binned data,
direct fitting has another feature: it heavily relies on
an assumption of a particular shape of the halo profile, which is typically taken as
the NFW profile. Halo profiles on average show systematic deviations
from the NFW approximation and are better approximated with the
Einasto profile \citep{Navarro04,Gao08,Navarro10}.  One may naively
expect that fitting a more accurate approximation would produce better
results on concentrations. This is not the case. There is a reason why
halo concentrations are not estimated with the Einasto profile:
results show large fluctuations in the concentration, which are due to
a smaller curvature of density profile $\rho(r)$ in the crucial range
of radii around the radius with the logarithmic slope -2. These
weaknesses of the direct fitting provide a motivation to look for
alternatives.

The idea is to start with the NFW profile.  It is uniquely defined by
the characteristic density $\rho_s$ and the characteristic radius
$r_s$. Instead of these 
two parameters one can use any two independent quantities to define
the profile.  For example, one can choose the virial radius and the
radius at which the circular velocity reaches half of its maximum
value as in \citet{Alam02}. Or one can use the radius containing 1/5
of virial mass and the virial mass \citep{Avila99}.  Recently
\citet{Klypin10} proposed to use a relation between the maximum
circular velocity $V_{\rm max}$, the virial mass and the virial radius
to estimate the halo concentration. We follow this idea and use a more
simple version with parameters $\Vmax$ and the circular velocity at
the virial radius:
\begin{equation}
   \V200=\left(\frac{G\M200}{\R200}\right)^{1/2}. 
\end{equation}
In this case the ratio $\Vmax/\V200$ of the maximum circular velocity
\Vmax~ to the virial velocity \V200~ is directly related with the halo
concentration and can easily be interpreted because larger ratios
imply larger concentration. The value of the halo concentration can be
estimated by assuming a shape of the density profile. However, this is
not required because the velocity ratio is a measure of the halo
concentration regardless of what profiles halos have.  For the case of
the NFW halo density profile, the $\Vmax/\V200$ velocity ratio is
given by the following relation:
\begin{equation}
 \frac{\Vmax}{\V200} = \left(\frac{0.216 \, c}{f(c)}\right)^{1/2}, 
\label{eq:eq1}
\end{equation}
where  $f(c)$ is 
\begin{equation}
 f(c) = \ln(1+c)-\frac{c}{(1+c)}. 
 \label{eq:eq2}
\end{equation}
Having the $\Vmax/\V200$ ratio for each halo, we find the halo
concentration $c$ by solving numerically eqs.~(\ref{eq:eq1} -
\ref{eq:eq2}). For convenience, Figure~\ref{fig:VratC} shows the
dependence of concentration $c$ on $\Vmax/\V200$ in a graphical
form. We also present results for three typical Einasto profiles
\citep{Navarro04,Gao08,Navarro10}:
\begin{equation}
\rho_{\rm Ein} = \rho_s\exp\left[-\frac{2}{\alpha}\left(x^\alpha-1\right)\right], 
           \quad x\equiv \frac{r}{r_s} \label{fig:Einasto}.
\end{equation}  

In practice we bin halos by their {\it maximum circular velocity}
$\Vmax$.  At a given redshift and for each circular velocity bin we
obtain the {\it median} virial velocity $\V200$. Then we compute $c$
by solving eqs.~(\ref{eq:eq1} - \ref{eq:eq2}).


\begin{figure}
\centering
\includegraphics[width=0.46\textwidth]{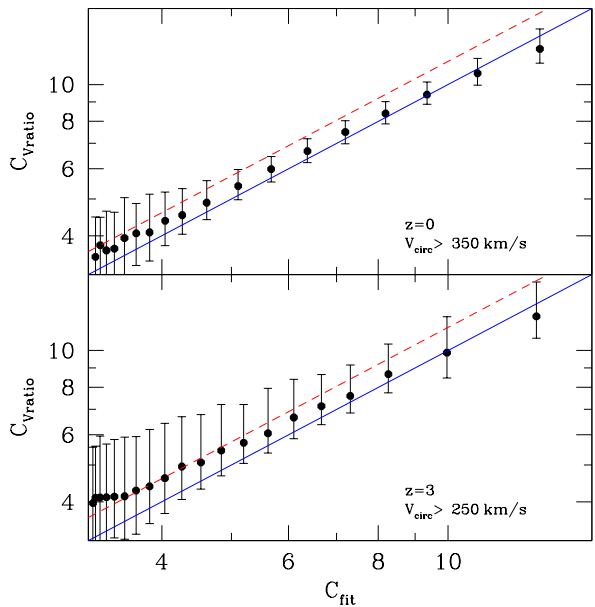}
\caption{Comparison of halo concentrations $c_{\rm  fit}$
  obtained by fitting the NFW profiles to halo densities with
  concentrations $c_{\rm Vratio}$ found using the ratios of the
  maximum circular velocity $V_{\rm max}$ to the halo velocity
  $V_{200}$. The most massive $\sim 10000$ halos in the Bolshoi
  simulation are used for the plot. The top (bottom) panel shows halos
  at $z=0$ ($z=3$). All halos are used regardless of their relaxation
  status.  Full lines present the relation $c_{\rm Vratio} =c_{\rm NFW
    fit}$.  The dashed lines show a 15\% offset between the two
  concentrations. Most of the median concentrations fall between those
  two lines. Error bars present 10 and 90 percent spread of the
  distribution of individual concentrations.}
\label{fig:ConcFit}
\end{figure}

 We made different tests of the accuracy of our algorithm.  We find
 quite a good agreement of our estimates of $c$ with the results of
 direct fitting of halo profiles by \citet{Neto07} with the deviations
 between the two methods being less than 5 percent for the whole range
 of masses in the MS simulation.
Figure~\ref{fig:Profiles} shows six examples of fitting density
profiles of large halos with $\M200\approx (1.5-3)\times
10^{14}\Msunh$ in the Bolshoi simulation. Each of the halos is
resolved with more than a million particles and the random noise
associated with finite number of particles is practically negligible.
We use 30 radial bins equally spaced in logarithm of radius with
$\Delta\log(r/\R200)=0.05$ up to $R=2\R200$. Plots show typical
examples of high and low concentration halos. Just as expected, for
halos with smooth density profiles (the bottom row of panels) the
accuracy of fits obtained with $\Vmax/\V200$ concentrations is very
good. Halos with large substructures and big radial fluctuations
(typically halos with low concentration; the top row of panels) result
in relatively poor fits. However, in these cases the direct fitting
does not produce much better approximations as demonstrated by the top
right panel.

Figure~\ref{fig:ConcFit} provides a more extensive comparison of
concentrations obtained with the direct fitting and with the
$\Vmax/\V200$ ratio method. For 
this plot we select all $\sim 7000$
halos in the Bolshoi simulation with $\Vmax > 350\,\kms$ at $z=0$ and
$\sim 10000$ halos with $\Vmax > 250\,\kms$ at $z=3$.  The halos with
concentration larger than $c\approx 5$ demonstrate a tight relation
between the two estimates of concentration: more than 90 percent of
halos fall between two lines: $c_{\rm Vratio}= c_{\rm fit}$ and $c_{\rm
  Vratio}= 1.15c_{\rm fit}$.  At smaller concentrations there is a small
systematic (5-15)\% offset between the two estimates.

These results demonstrate that both the direct fitting of density
profiles and the method based on the $\Vmax/\V200$ ratio give similar
concentrations.  There are the small systematic deviations, which for
relaxed halos or for halos with concentrations $c>5$ are less than a few
percent. For halos with small concentrations the deviations are larger,
but are still less than $\sim 15\%$. We ignore these differences and use
concentrations based on the  $\Vmax/\V200$ ratios.

 \section{Evolution of the distinct halo concentrations with redshift:
   results of simulations}

\begin{figure}
\centering
\includegraphics[width=0.46\textwidth]{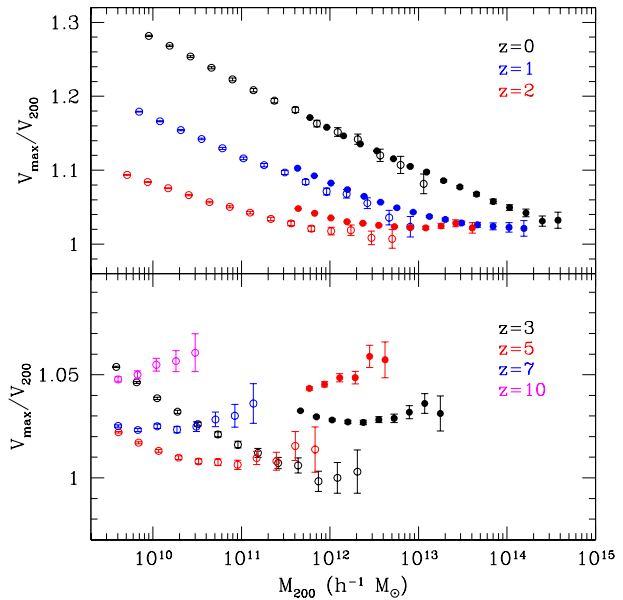}
\caption{The ratio $\Vmax/\V200$ of the maximum circular velocity to
  the virial velocity as a function of mass $\M200$ for distinct halos
  at different redshifts for MS-I (filled symbols) and MS-II (open symbols) simulations. Error bars are
  statistical uncertainties.  The MS-I and MS-II simulations agree quite well at $z=0$.
  At higher redshifts there are noticeable differences between MS-I and MS-II.}
\label{fig:fig3a}
\end{figure}
\begin{figure}
\centering
\includegraphics[width=0.46\textwidth]{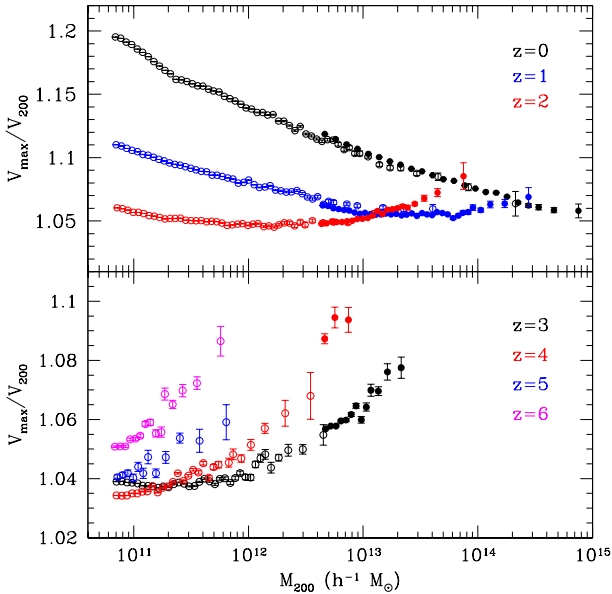}
\caption{The same as Figure~\ref{fig:fig3a} but for Bolshoi (open
  symbols) and MultiDark (filled symbols) simulations. Both
  simulations show remarkable agreement at all masses and redshifts.}
\label{fig:fig3b}
\end{figure}

\begin{figure}
\centering
\includegraphics[width=0.48\textwidth]{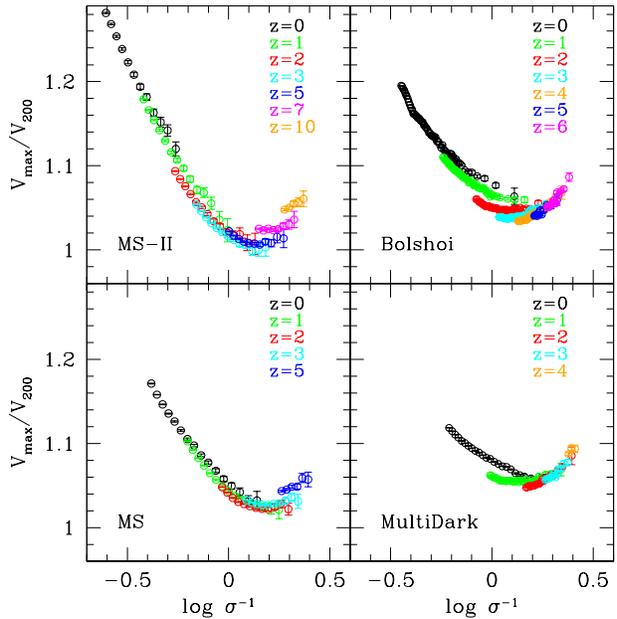}
\caption{The ratio $\Vmax/\V200$ as a function of $\log \,
  \sigma^{-1}$. In this plot halos mass increases from left to
  right. The large variations in shapes and amplitudes of
  $\Vmax/\V200 - M$ seen in Figures~\ref{fig:fig3a}-\ref{fig:fig3b} are
  replaced with much tighter relations $\Vmax/\V200 - \log \,
  \sigma^{-1} $. All simulations show the same
  pattern: $\Vmax/\V200$ has a U-shape with an upturn at large masses.}
\label{fig:fig4}
\end{figure}

Our analysis of halo concentrations in different simulations starts
with the simpler statistics of $\Vmax/\V200$. It is directly related
with the concentration $c$, but $\Vmax/\V200$ does not require an
assumption of a particular density profile. These raw data are
presented in Figures~\ref{fig:fig3a}-\ref{fig:fig3b} for all
simulations discussed in section~2.  Here we consider only halos with
more than 500 particles for our analysis. Error bars correspond to the
statistical uncertainty in the determination of the
$\Vmax/\V200$ median velocity ratio.  Halos are binned by $\Vmax$, and the
average $\M200$ for each velocity bin is 
the horizontal axis.

At $z=0$, the agreement between both Millennium simulations is very
good in the range of masses where they overlap. This is expected if
the two simulations faithfully present results of the cosmological
model and are not affected by finite box sizes or any other numerical
issues.  Yet, at higher redshifts there are notable differences
between MS-I and MS-II (see Figure~\ref{fig:fig3a}). At a fixed halo mass $\M200$, data from MS-I
yield systematically larger $\Vmax/\V200$ velocity ratios, increasing
with redshift up to a $3-4\%$ difference. For example, at $z=3$ we
find $\sim 3\%$ difference for $~10^{12}\Msunh$. This will translate
to a difference of about $20-40\%$ in terms of halo concentration
according to eq.~(\ref{eq:eq1}).

Regardless of this discrepancy between the two Millennium simulations, we
do clearly see an upturn of the velocity ratios, and, hence, halo
concentrations 
with increasing halo mass
at higher redshifts.  The same is observed in the
Bolshoi and MultiDark simulations (see Figure~\ref{fig:fig3b}). We
conclude that concentrations of distinct halos do increase with halo
mass at higher redshifts as previously reported by \citet{Klypin10}.
In spite of the qualitative agreement, there are differences between
Bolshoi/MultiDark and the Millennium-I/II simulations. Those are
expected and are due mainly to the different cosmological parameters
used in the simulations.

The reason for the discrepancy between
Millennium simulations found in the $\Vmax/\V200$ velocity ratios at
high redshifts is not clear. It might be due to differences in mass and force
resolution.  If that were true, one may expect that a lower resolution
MS-I simulation should have a lower 
concentration. However, this is not
the case: MS-I has larger concentration for overlapping masses.  An
important aspect that could cause this discrepancy is the fact that
 substructure within the dominant (distinct) halo is not taken
into account when $\Vmax$ is computed. Certainly this could make
differences up to the level of few percent as observed. The 
MultiDark simulation is helping us to clarify this issue because
MultiDark and Bolshoi have very different resolutions.  In this case,
we do see in Figure~\ref{fig:fig3b} a good agreement with the Bolshoi
simulation at all redshifts in the mass interval where both
simulations overlap. The BDM halo finder does include substructures
within distinct halos to compute $\Vmax$.

So far we were following the traditional path of studying
concentrations by expressing results for $c$ as functions of halo mass
at different redshifts $c =c(M,z)$ \citep[e.g.,][]{Bullock01,Eke01,
  Zhao03a, Maccio08}. As many other groups find, results for $c(M,z)$
appear to be not simple: the concentration evolves in a complicated
way.  In addition to previously found effects, at high redshifts the
amplitude starts to increase, which makes the whole $c(M,z)$ function
very complex.

One of reasons why $c(M,z)$ may {\it look} too complicated is because
``wrong'' physical quantities $M$ and $z$ are used. The situation may
be compared to the studies of the halo mass function. Once a more
physical variable $\sigma(M,z)$ was used \citep[see
e.g.][]{Jenkins01,ST02,Warren06,Tinker08}, the results became more transparent.  In
turn, this simplification allowed the development of accurate
approximations such as that presented by
eqs.~(\ref{eq:dndm}-\ref{eq:sigma}). We will follow the same direction
when we study the halo concentration. In particular, as the main
variable we choose $\log\, \sigma^{-1}(M,z)$ - a quantity that is
widely used in the mass function analysis.  Using this variable helps
to remove most of the evolution of the concentration with redshift, helps
to understand the differences due to the cosmological parameters and the
matter power spectrum, and allows a more clear comparison between
different simulations.

\begin{figure}
\centering
\includegraphics[width=0.48\textwidth]{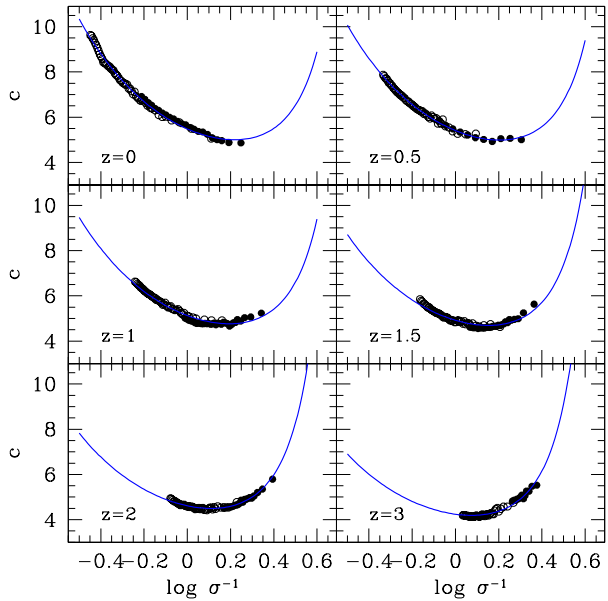}
\caption{Concentration-$\sigma(M)$ median relation for distinct halos
  in the Bolshoi (open symbols) and MultiDark (filled symbols)
  simulations for different redshifts. Solid lines are predictions
  from our analytical model given by
  eqs.(\ref{eq:eq3}-~\ref{eq:pars}). The data at different redshifts
  look very similar, but actually have a slight evolution: the minimum
  of $c$ becomes smaller and it is reached at smaller $\log \, \sigma^{-1}$}
\label{fig:fig6}
\end{figure}

Figure ~\ref{fig:fig4} shows the $\Vmax/\V200$ velocity ratio as a function of
$\log \, \sigma^{-1}$ of distinct halos at different redshifts for
both Millennium simulations (left panels) and Bolshoi/MultiDark (right
panels). Large volume simulations MS-I \& MultiDark provide good
statistics at the high-mass end ($\log \, \sigma^{-1} > 0$) and the
smaller boxes, with much higher numerical resolution (MS-II and
Bolshoi), help to reach small values of $\log \, \sigma^{-1} < 0$
(small masses). Figure ~\ref{fig:fig4} also shows that 
when expressed in terms of the variable $\log \, \sigma^{-1}$, the concentration becomes a
well-behaved function with an U shape: at low $\log
\, \sigma^{-1}$ it decreases with $\log
\, \sigma^{-1}$. Then it reaches a minimum  at a well defined scale $\log \, \sigma^{-1}
\approx 0.15$ ($\sigma\approx 0.707)$, and at larger $\log \, \sigma^{-1}$ it begins to increase.

\begin{figure}
\centering
\includegraphics[width=0.45\textwidth]{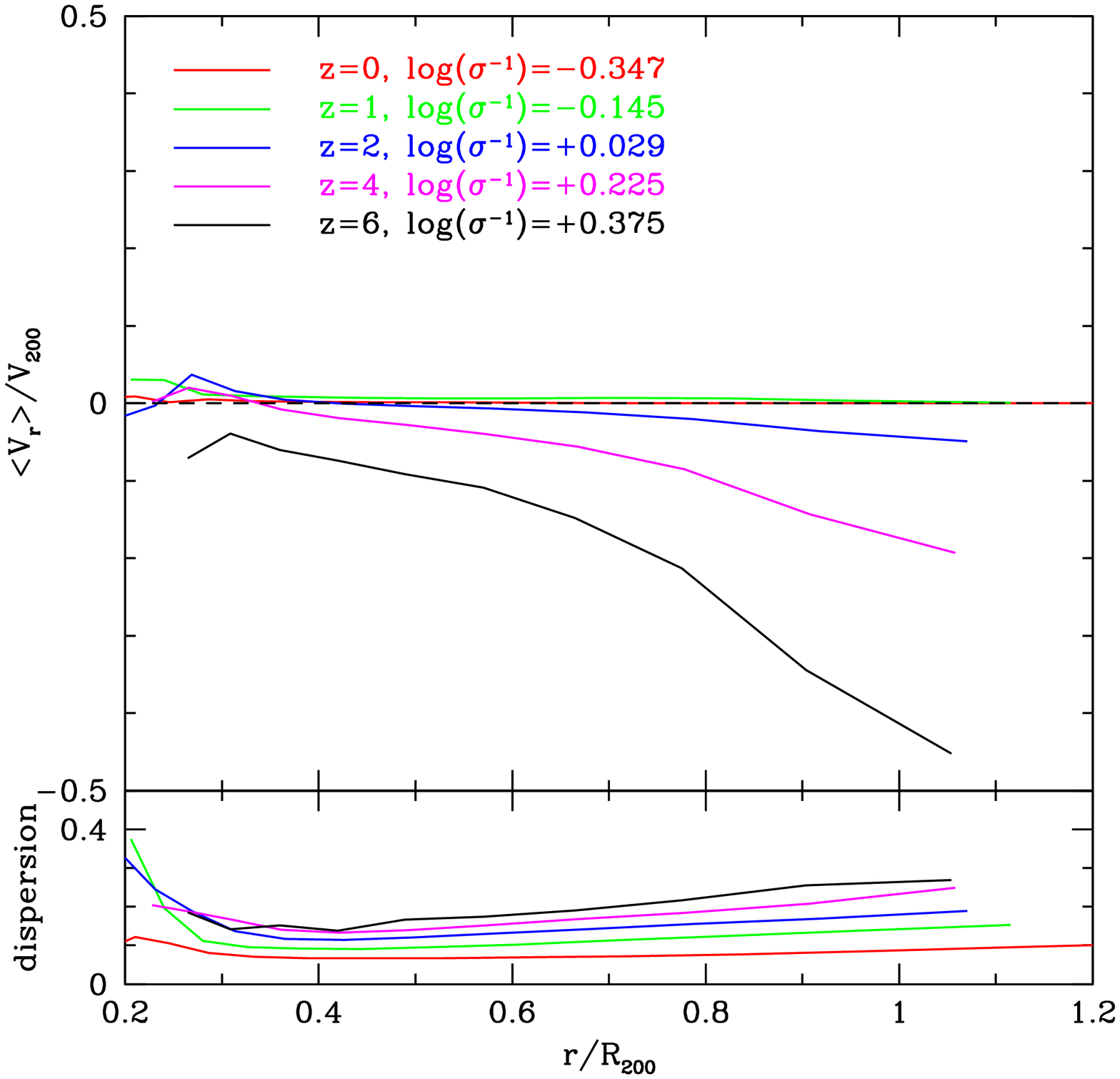}
\includegraphics[width=0.45\textwidth]{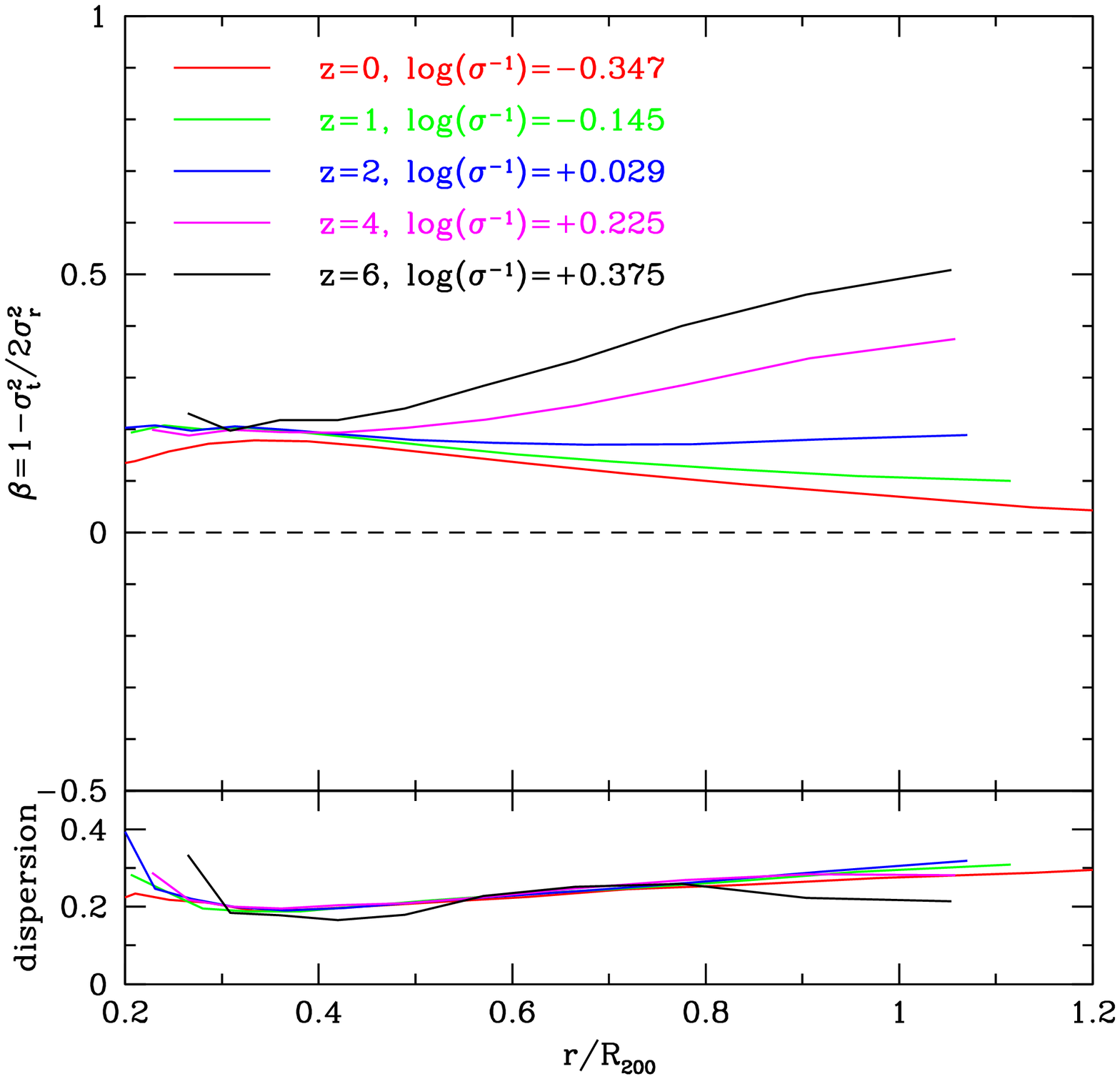}
\caption{Average radial velocity $\Vrad$ (top panels) and velocity
  anisotropy $\beta$ (bottom panels) profiles for distinct halos
  selected from the Bolshoi simulation in different bins of $\log \,
  \sigma^{-1}$ and redshifts indicated in the plots. Halos with large
  $\log \, \sigma^{-1}$ (most massive halos) clearly show
  preferentially radial orbits and overall infall pattern at large
  distances from the halo center. Less massive halos ($\log \,
  \sigma^{-1}<0$) on average do not have infall velocities and their
  orbits in the peripheral regions tend to be isotropic while retaining
  slightly radial velocities in the central regions.}
\label{fig:fig5}
\end{figure}    
    
In spite of the fact that the rescaling from $M$ to $\log \,
\sigma^{-1}$ significantly reduces deviations in  $\Vmax/\V200$
(and, thus, in concentration), it does not completely remove all the deviations.
There is still some residual dependence on cosmology: there are
differences between the Millennium and the Bolshoi/MultiDark
simulations. In addition, there is a weak evolution with
redshift, which is more pronounced in high resolution runs Bolshoi and
MS-II

As discussed above, at high redshifts there is a discrepancy between
MS-I and MS-II simulations resulting in a $~20-40\%$ difference in
concentrations for a given halo mass $M$ between the two
simulations. At the same time, the overlap between Bolshoi and
MultiDark is good at all redshifts. This is the reason we use Bolshoi
and MultiDark to study the weak
 evolution  of the $c-\log \, \sigma^{-1}$ relation
with the redshift.
The time evolution of the concentration
can be described as a decrease of the
minimum value of the U-shaped concentration curves with increasing
redshift (from $~5.1$ at $z=0$ down to $~3.7$ at $z=4$), and a slight
shift of the position of the minimum to smaller values of $\log \,
\sigma^{-1}$, i.e. from $~0.2$ at $z=0$ to $~0.05$ at $z=4$.  Figure
~\ref{fig:fig6} shows details of this weak evolution.

The nature of the upturn in the halo concentrations is an interesting
problem on its own. We do not make detailed analysis of halos in the
upturn, but we test some simple hypotheses. One may think that the
upturn is due to non-equilibrium effects. Indeed, the halos in the
upturn are the largest halos at any given moment, and those halos are
known to grow  very fast. Results presented in the Appendix
clearly show that out-of-equilibrium effects do not provide an
explanation for the upturn: selecting relaxed halos only increases the
magnitude of the upturn. However, the fast growth of the halos in the
upturn does change the structure of the most massive halos (even
those that are ``relaxed'').

Figure~\ref{fig:fig5} shows average radial velocity profiles $\Vrad$
and velocity anisotropy $\beta$ profiles for distinct halos in the
Bolshoi simulation for different bins of $\log \, \sigma^{-1}$. These
statistics can be used to indicate the dynamical state of 
halos that lie at different locations in the
$\Vmax/\V200$--$\sigma(M)$ median relations displayed in Figure
~\ref{fig:fig4}. Average halos in the high-mass end (i.e. $\log \,
\sigma^{-1} > 0$) clearly show signatures of infall given by large
negative radial velocities. This is more clearly seen when expressed
in terms of $\Vrad$, but is also present in the velocity anisotropy
$\beta=1-\sigma_t^2/2\sigma_r^2$, which tends to be larger at large
values of $\log\sigma^{-1}$. These results indicate that orbits are
preferentially radial for halos in the upturn part of the $c-M$
relation.  The situation changes for halos with smaller $\log \,
\sigma^{-1}$. They do not show an infall pattern and their velocity anisotropy
is significantly smaller.  Density profiles are less sensitive to
these effects. The halos are ``normal'': on average they are well
described by NFW for all $\log \sigma^{-1}$ bins. The only indication
that something is not normal is the large halo concentration.


 
 \section{The $c-\sigma(M)$ relation: an analytical model for the halo
   mass--concentration relation}

It is then desirable to provide for the $\LCDM$ cosmology an
approximation that describes the dependence of halo concentration $c$
on halo mass and redshift. 
However, approximations which use mass and
redshift as the main variables are prone to a severe problem: lack of
scalability.  An approximation found for one set of cosmological
parameters and for one particular redshift is not applicable to another
cosmology or  redshift. 
Just as with the mass function, we find a dramatic improvement of the
accuracy of the approximations for the halo concentration once we use
$\sigma(M,z)$.  Following the same line of ideas, we use a new ``time
variable'' $x$ defined below. The motivation comes from the linear
growth rate $D$ of fluctuations in the $\LCDM$ cosmology. If it is
normalized to 
be unity at $z=0$, then

\begin{eqnarray}
D(a) &=& \frac{5}{2}\left( \frac{\Omega_{\rm m,0}}{\Omega_{\Lambda,0}} \right)^{1/3}
                            \frac{\sqrt{1+x^3}}{x^{3/2}}
                              \int_0^x\frac{x^{3/2}dx}{[1+x^3]^{3/2}},
\label{eq:eq12}\\
  x &\equiv& \left( \frac{\Omega_{\Lambda,0}} {\Omega_{\rm m,0}} \right)^{1/3}a, \quad a \equiv (1+z)^{-1},
\label{eq:x}
\end{eqnarray}
where $\Omega_{\rm m,0}$ and $\Omega_{\Lambda,0}$ are the matter and
cosmological constant density contributions at $z=0$. Note that when
$D$ is written in this form, there is no explicit dependence on the
redshift $z$: the time dependence goes through the variable $x$.

If the concentration $c$ depended only on $\sigma$, we
would had a universal function $c(\sigma)$.  Similar to the mass
function, we find that concentration is not universal and has some
dependance on the redshift. We attribute this residual redshift
dependence of $c(\sigma,a)$ mostly to the change in the growth rate of
fluctuations $D(a)$ related to the change of matter density parameter
$\Omega_{\rm m}(a)$.  Another possible factor is the change with the
redshift of the slope of the power spectrum $P(k)$ at a given
$\sigma$. We find that corrections due to the growth rate factor
produce sufficiently accurate fits and we do not study here the effect due to
change in the $P(k)$ slope. 

\begin{figure}
\centering
\includegraphics[width=0.48\textwidth]{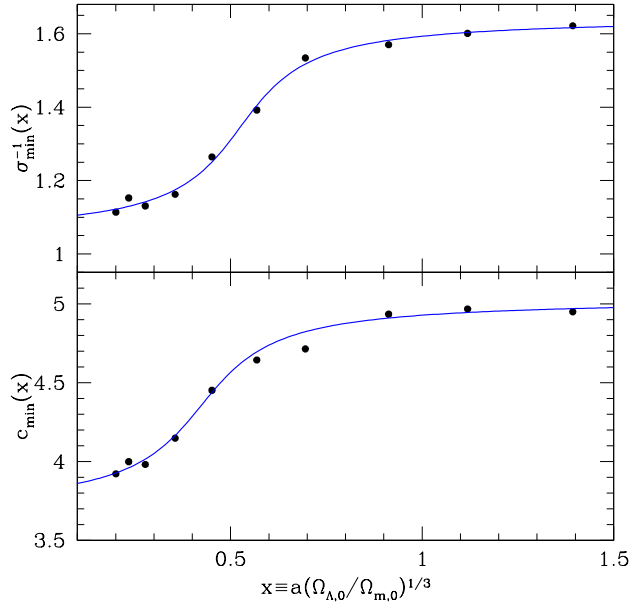}
\caption{Dependence  of the halo minimum concentration
  $c_{\rm min}$ (bottom panel) and the value of $\sigma^{-1}_{\rm
    min}$ (top panel) on $x$ for distinct halos in the Bolshoi 
  and MultiDark  simulations. Solid lines are the
  approximations given in eqs.(\ref{eq:eq8}-~\ref{eq:eq9}).}
\label{fig:fig7}
\end{figure}

Because all our approximations use only $\sigma$ and $x$, we {\it expect}
that they are applicable also for different redshifts and different
cosmological parameters.  Fitting of all available data from Bolshoi
and MultiDark simulations gives the following approximation for halo
concentration:
\begin{eqnarray}
  c(M,z)  &=& B_0(x)\, \mathcal{C}(\sigma'), \label{eq:eq3}\\
  \sigma'&=& B_1(x)\, \sigma(M,x), \label{eq:eq4}\\
  \mathcal{C}(\sigma') &=& A \, \left[\left(\frac{\sigma'}{b}\right)^c+1\right] 
                  \exp\left(\frac{d}{\sigma'^2}\right), \label{eq:eq5}
\end{eqnarray}
where 
\begin{equation}
 A=2.881,\, b=1.257,\, c=1.022,\, d=0.060
\label{eq:pars}
\end{equation}
The  parameterization form of the function was motivated
by the Sheth-Tormen \citep{ST02} approximation for the halo mass function.
Here, the functions $B_0(x)$ and $B_1(x)$ are defined in such a way that
they are equal to unity at $z=0$ for the WMAP-5 parameters of the
Bolshoi and MultiDark simulations. Thus, the function
$\mathcal{C}(\sigma')$ is the concentration at $z=0$ for this
cosmological model. We find the following approximations for $B_0(x)$ and $B_1(x)$:

\begin{equation}
B_0(x) = \frac{c_{\rm min}(x)}{c_{\rm min}(1.393)}, \quad
B_1(x) = \frac{\sigma^{-1}_{\rm min}(x)}{\sigma^{-1}_{\rm min}(1.393)}, \label{eq:eq7}
\end{equation}
where $c_{\rm min}$ and $\sigma^{-1}_{\rm min}$ define the minimum of the
halo concentrations and the value of $\sigma$ at the minimum:
\begin{eqnarray}
c_{\rm min}(x) = c_0+ (c_1- c_0)
    \left[\frac{1}{\pi}\arctan\left[\alpha(x-x_0)\right] + \frac{1}{2} \right] \label{eq:eq8} \\
\sigma^{-1}_{\rm min}(x)= \sigma^{-1}_0+ (\sigma^{-1}_1-\sigma^{-1}_0)
     \left[\frac{1}{\pi}\arctan\left[\beta(x-x_1)\right] +\frac{1}{2} \right],\label{eq:eq9}
\end{eqnarray}
where 
\begin{equation}
c_0 = 3.681,\, c_1 = 5.033,\, \alpha = 6.948,\, x_0 = 0.424,
\label{eq:pars2}
\end{equation}
 and
\begin{equation}
 \sigma^{-1}_0 = 1.047,\,  \sigma^{-1}_1 = 1.646,\, \beta = 7.386,\, x_1 = 0.526.
\label{eq:pars3}
\end{equation} 
Accurate approximations for the $rms$ density fluctuation
$\sigma(M,a)$ for the cosmological parameters of the Bolshoi/MultiDark
simulations are given in \citet{Klypin10} and for convenience are
reproduced here:

\begin{eqnarray}
\sigma(M,a) &=& D(a) \, \frac{16.9 \, y^{0.41}}{1+1.102 \, y^{0.20}+6.22 \, y^{0.333}},\label{eq:eq11} \\
    y &\equiv& \left[\frac{M}{10^{12}\Msunh} \right]^{-1}. \nonumber 
\end{eqnarray}

\begin{figure}
\centering
\includegraphics[width=0.48\textwidth]{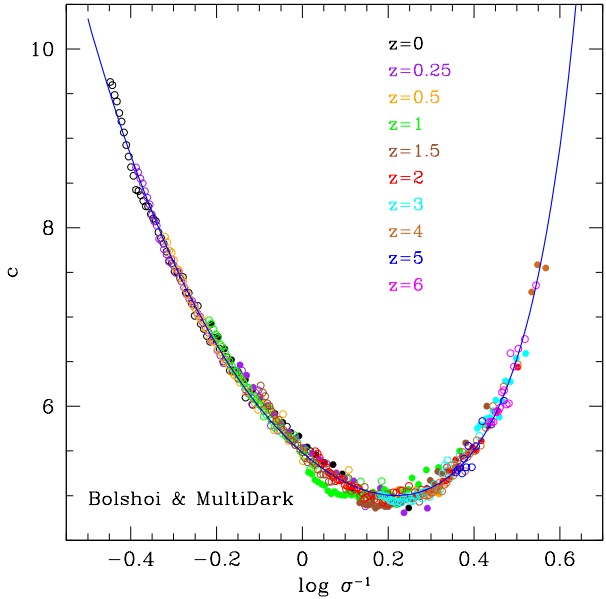}
\caption{Dependence of halo concentration $c$ on $\log\,\sigma^{-1}$
  after rescaling all the results of Bolshoi and MultiDark
  simulations to $z=0$. The plot
  shows a tight intrinsic correlation of $\mathcal{C}$ on $\sigma'$.
}
\label{fig:fig8}
\end{figure}

Figure~\ref{fig:fig7} shows the evolution of $c_{\rm min}$ and
$\sigma^{-1}_{\rm min}$ with ``time'' $x$, and presents the
approximations given in eqs.(\ref{eq:eq8}-\ref{eq:eq9}). The evolution
is clearly related with the transition from the matter dominated
period ($\Omega_{\rm m}(a)\approx 1$, $x<0.3$) to the
$\Lambda$-dominated one with $x>0.7$.  Approximations for the halo
concentration are presented in Figure~\ref{fig:fig6} for some
redshifts.  The parameters $A,b,c,d$ of the $\mathcal{C}(\sigma')$
relation are determined from the best fit to the
concentration--$\sigma(M)$ Bolshoi/MultiDark data at all redshifts.

\medskip
Here is a step-by-step description how to estimate halo concentration:
\begin{itemize}
  \item For given mass $M$ and  $a=1/(1+z)$ find $x$, $D(a)$, and $\sigma(M,a)$
        using eqs.~(\ref{eq:x}, \ref{eq:eq12}, \ref{eq:eq5} or \ref{eq:eq11})
  \item Use eq.~(\ref{eq:eq7}) to find parameters $B_0$ and $B_1$.
  \item Use eqs.~(\ref{eq:eq4}-\ref{eq:eq5}) to find  $\sigma'$ and $\mathcal{C}$
   \item Use eq.~(\ref{eq:eq3}) to find halo concentration $c(M,z)$.
\end{itemize}
  
We present the final results and approximations in two different
forms.  Functions $B_0$ and $B_1$ can be used to find values of
$\mathcal{C}$ and $\sigma'$, which is effectively the same as
rescaling concentrations $c(\sigma,x)$ measured in simulations to the
same redshift $z=0$. Figure~\ref{fig:fig8} shows results of
simulations rescaled in this way.  The U-shape of
$\mathcal{C}(\sigma')$ is clearly seen. The $\mathcal{C}(\sigma')$
function to some degree plays the same role for concentrations as the
function $f(\sigma)$ for the mass function in
eqs.(\ref{eq:dndm}-\ref{eq:fsigma}). It tells us that there is little
evolution in the dependence of concentration with mass once intrinsic
scalings (e.g., $x$ instead of expansion parameter) are taken into
account.

\begin{figure}
\centering
\includegraphics[width=0.49\textwidth]{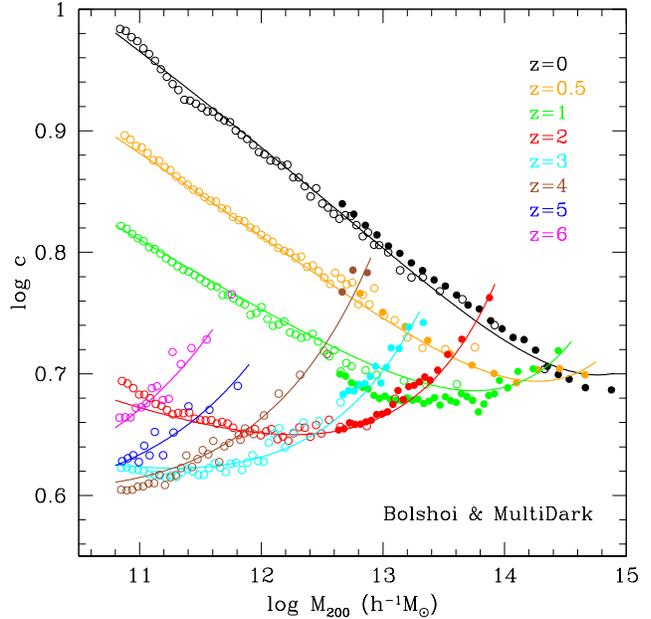}
\caption{Halo mass--concentration relation of distinct halos at
  different redshifts in the Bolshoi (open symbols) and MultiDark
  (filled symbols) simulations is compared with analytical
  approximation eqs.(\ref{eq:eq3}-\ref{eq:eq5} (curves)). The errors of the
  approximation are less than a few percent.}
\label{fig:fig9}
\end{figure}

Another way of showing the approximations is simply plot
eqs. (\ref{eq:eq3}-\ref{eq:eq5}) for different redshifts and compare
the results with the median concentration - mass relation  in our simulations. This comparison is
presented in Figure~\ref{fig:fig9}. It shows that the errors of the
approximation are just a few percent for the whole span of masses and
redshifts.

\begin{figure}
\centering
\includegraphics[width=0.46\textwidth]{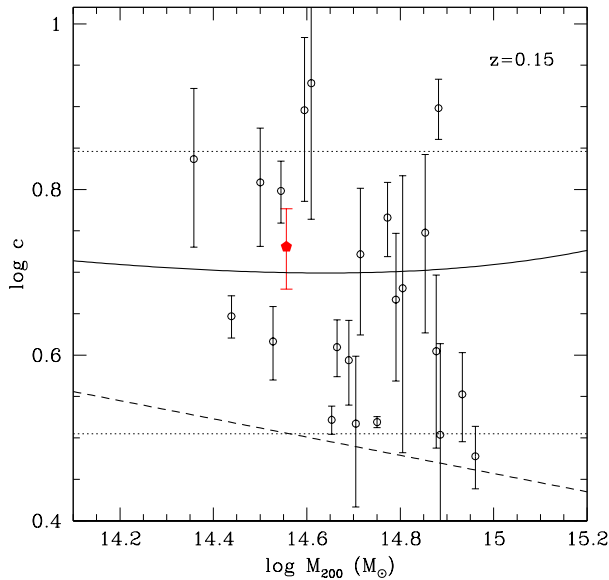}
\caption{Comparison of observed cluster concentrations (data points
  with error bars) with the prediction of our model for median halo
  concentration of cluster-size halos (full curve). Dotted lines show 10\% and 90\%
  percentiles. 
  Open circles show results for
  X-ray luminous galaxy clusters observed with XMM-Newton in the
  redshift range 0.1-0.3 \citep{Ettori10}. The pentagon presents
  galaxy kinematic estimate for relaxed clusters by \citet{WL10}. The
  dashed curve shows prediction by \citet{Maccio08}, which significantly 
  underestimates the concentrations of clusters.
  \label{fig:Clusters}}
\end{figure}

\section{Summary and Conclusions}

We study the halo concentrations in the $\LCDM$ cosmology, from the
present up to redshift ten, over a large range of scales going from
halos similar to those hosting dwarf galaxies to massive galaxy
clusters, i.e. halo maximum circular velocities ranging from 25 to
1800 $\kms$ (about six orders of magnitude in mass), using
cosmological simulations with high mass resolution over a large
volume. The results presented in this paper are based on the Bolshoi,
MultiDark, and Millennium-I and II simulations. There is a good consistency
among the different simulation data sets despite the
different codes, numerical algorithms, and halo/subhalo finders used
in our analysis.

The approximations given here for the evolution of the halo
concentration constitute the state-of-the-art of our current knowledge
of this basic property of dark matter halos found in N-body $\LCDM$
simulations. Naturally, new simulations that improve mass resolution in
even larger volumes are needed to face the challenges imposed by
current and future observational programs. Our analysis can also be
useful for comparison with analytical works that aim to understand the
statistics and structural properties of dark matter halos in the
standard $\LCDM$ cosmology.

It is interesting to compare these results with other simulations and
models in the literature. \citet[][]{Zhao03a,Zhao09} were the first to
find that the concentration flattens at large masses and at high
redshifts. Actually, figure 15 in \citet[][]{Zhao09} also shows an
upturn in concentration at $z = 4$. However, the authors do not even
mention it in the text. In addition, their model prediction of the
halo mass--concentration relation, based on the halo mass accretion
histories, failed to reproduce the upturn behavior of the
concentration with increasing mass.

For $\Mvir=$ $10^{12}\Msunh$ in the MS-II and Aquarius simulations
\citet{BK09} give concentration of $c_{\rm vir}=12.9$, which is
1.3 times larger than what we get from Bolshoi.  Most of the
differences are likely due to the larger amplitude of cosmological
fluctuations in MS simulations because of the combination of a larger
$\sigma_8$ and a steeper spectrum of fluctuations.  On larger masses
\citet{Neto07} give the following approximation for all halos for MS
cosmological parameters: $c_{200} =
7.75(M_{200}/10^{12}\Msunh)^{-0.11}$.  Thus, the MS-I has a small
($\sim $10\%) difference in $c_{200}$ as compared with our results for
$\Mvir=$ $10^{14}-10^{15}\Msunh$.

If we use the same selection conditions (all halos selected by mass)
and use the same cosmological model as in \citet{Maccio08}, then the
concentrations of Milky-Way-size halos ($\M200 =10^{12}\Msunh$ in
\citet{Maccio08} are 10\% lower than what we find -- a reasonable
agreement.  However, our results are in contradiction with those of
\citet{Maccio08} when we consider clusters of galaxies. For example,
for $\M200 =5\times 10^{14}\Msunh$ we find $c_{200} =4.6$ while
approximations in \citet{Maccio08} give a substantially lower value of
$c_{200} =3.1$ -- a 50\% smaller value.

In Figure~\ref{fig:Clusters} we compare predictions of our model
eqs.(\ref{eq:eq3}-\ref{eq:eq5}) for the median mass--concentration
relation for galaxy clusters (solid line) with observational estimates
obtained from state-of-the-art estimates based on X-ray
\citep{Ettori10} and kinematic \citep{WL10} data. The model has been
computed at redshift $z=0.15$, the median of the redshift distribution
of the X-ray sample. Open circles are c--$\M200$ measurements for the
sample of 23 X-ray luminous galaxy clusters observed with XMM-Newton
in the redshift range 0.1-0.3 selected from Table 2 in
\citet{Ettori10}. We adopted for each cluster an average concentration
obtained from both concentration measurements using two different
techniques applied by the authors to recover the gas and dark matter
profiles. We rejected those clusters where their concentration
estimates differs more than
$30\%$ between both mass reconstruction techniques. The
mass-concentration mean estimate from the kinematic analysis of the
combined sample of 41 nearby clusters is also shown as a pentagon
symbol \citep{WL10}. The original estimate of \citet{WL10} has been
corrected to our overdensity definition. Our model prediction is in
good agreement with the mass--concentration observational
measurements. Yet, the median halo mass-concentration relation from
Maccio et al. 2008 (dashed line) produces by far lower estimates of the concentration for a
given value of the halo mass as compared to the observational data.

\medskip
Our main results can be summarized as follows:

\begin{itemize}

\item Our study of the evolution of the $\Vmax/\V200$ velocity ratio
  (which is a measure of concentration) as a function of halo mass $\M200$ has
  confirmed that the halo mass--concentration relation shows a
  novel feature at high redshifts: a flattening and an upturn at the
  high-mass end.
\item When expressed in terms of the variable $\log \, \sigma^{-1}$,
  the halo concentration is a well-behaved function with a U-shaped
  trend: at small masses the concentration first declines, reaches a
  minimum at $\log \, \sigma^{-1} \sim 0.15$, and then increases again
  at larger masses. The $c-\sigma$ relation is much narrower than the
  traditional $c(M,z)$ relation, but it is not exactly universal:
  there are some small dependences with redshift and cosmology.
\item The median concentration--$\sigma(\M200)$ relation can be
  accurately parameterized by eqs.(\ref{eq:eq3}-\ref{eq:eq5}). This relation provides
  an analytical model of the halo mass--concentration median relation
  $c(\M200)$ that reproduces all the relevant features, namely the
  decline of concentration with mass, and its flattening and upturn
  at high redshift/mass.
\item Our estimates for concentration of cluster-size halos are
  compatible with the recent observational results and are
  substantially -- a factor of 1.5 -- larger than in \citet{Maccio08}.

\end{itemize}

\section*{Acknowledgments}

We are grateful to Gerard Lemson for help with the Millennium
databases. The Millennium and Millennium-II
simulation databases used in this paper and the web application
providing online access to them were constructed as part of the
activities of the German Astrophysical Virtual
Observatory.  
We thank Michael Boylan-Kolchin, Simon White, Gerard Lemson,
and Raul Angulo for double checking some of our results and for fruitful
comments on the manuscript. We also thank Shaun Cole, Adrian Jenkins,
Radek Wojtak, Miguel Sanchez-Conde, and Andrey Kravtsov for useful
discussions. We acknowledge the support of the NSF grants to NMSU 
and UCSC, and
the support of the Spanish MICINN Consolider-Ingenio 2010 Programme
under grant MULTIDARK CSD2009-00064. The Bolshoi and MultiDark (BigBolshoi) simulations were run on the
Pleiades supercomputer at NASA Ames Research Center.
F. P. thanks the hospitality of the Institute for
Computational Cosmology at Durham University and MPA at Garching where part of this work
was done.

\section{APPENDIX: Effects of halo selection}

Selection of halos has some impact on estimates of halo
concentrations. There are different ways to select halos and each
selection condition has its own effect.  One of the possibilities is
to select quiet halos from a population of all halos. 
 Halos can be
selected by mass or by the maximum circular
velocity $\Vmax$, as we do in this paper. Halo radius may also be defined differently, which
substantially changes the halo concentration. Here we consider and discuss 
the impact
of different selection effects.

\begin{figure}
\centering
\includegraphics[width=0.46\textwidth]{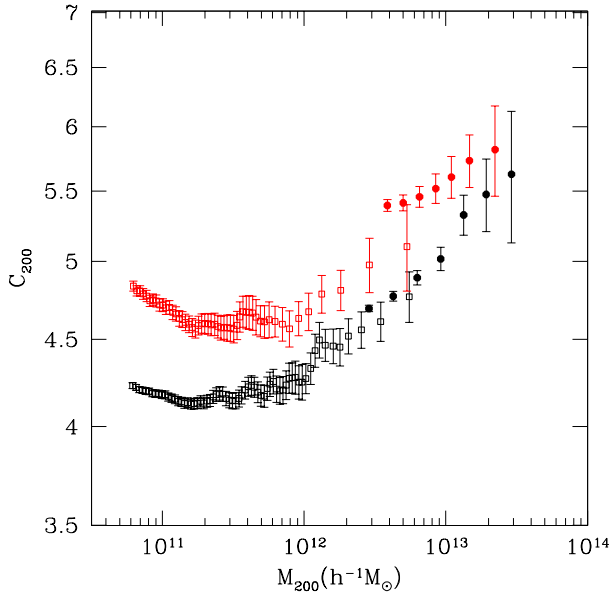}
\caption{
  Effects of relaxation on halo concentration at redshift $z=2$. Top
  symbols show only relaxed halos. Concentrations of all halos are
  shown by bottom symbols. Relaxed halos are more concentrated and,
  just as all halos, show an upturn at large masses.  Open symbols
  show results from the Bolshoi simulation; filled symbols are for the
  Multidark simulation. Halos are binned by $\Vmax$. The average halo mass $\M200$
  for each velocity bin is shown on the horizontal axis.}
\label{fig:EffectsB}
\end{figure}

\begin{figure*}
\centering
\includegraphics[width=0.46\textwidth]{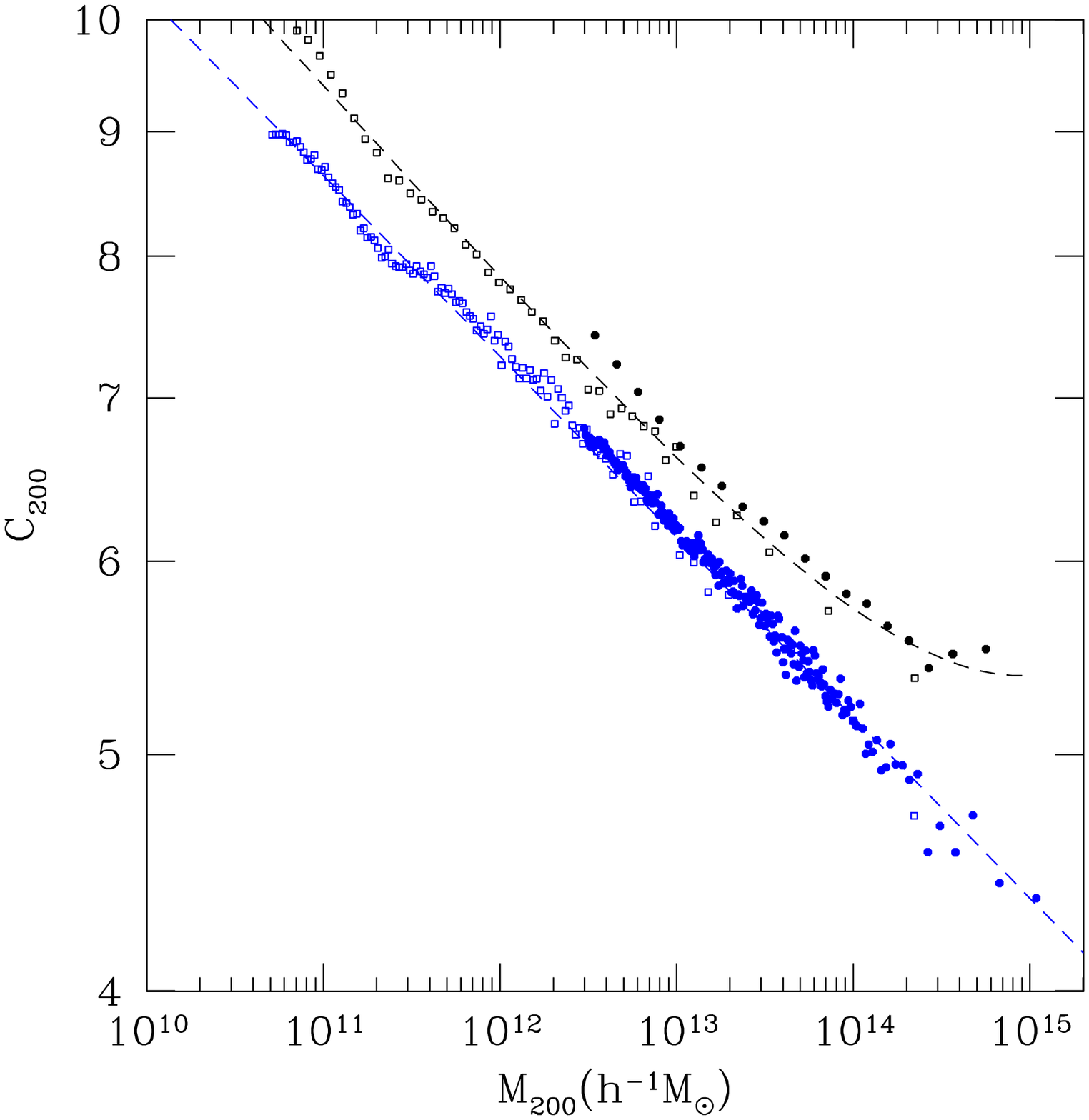}
\includegraphics[width=0.46\textwidth]{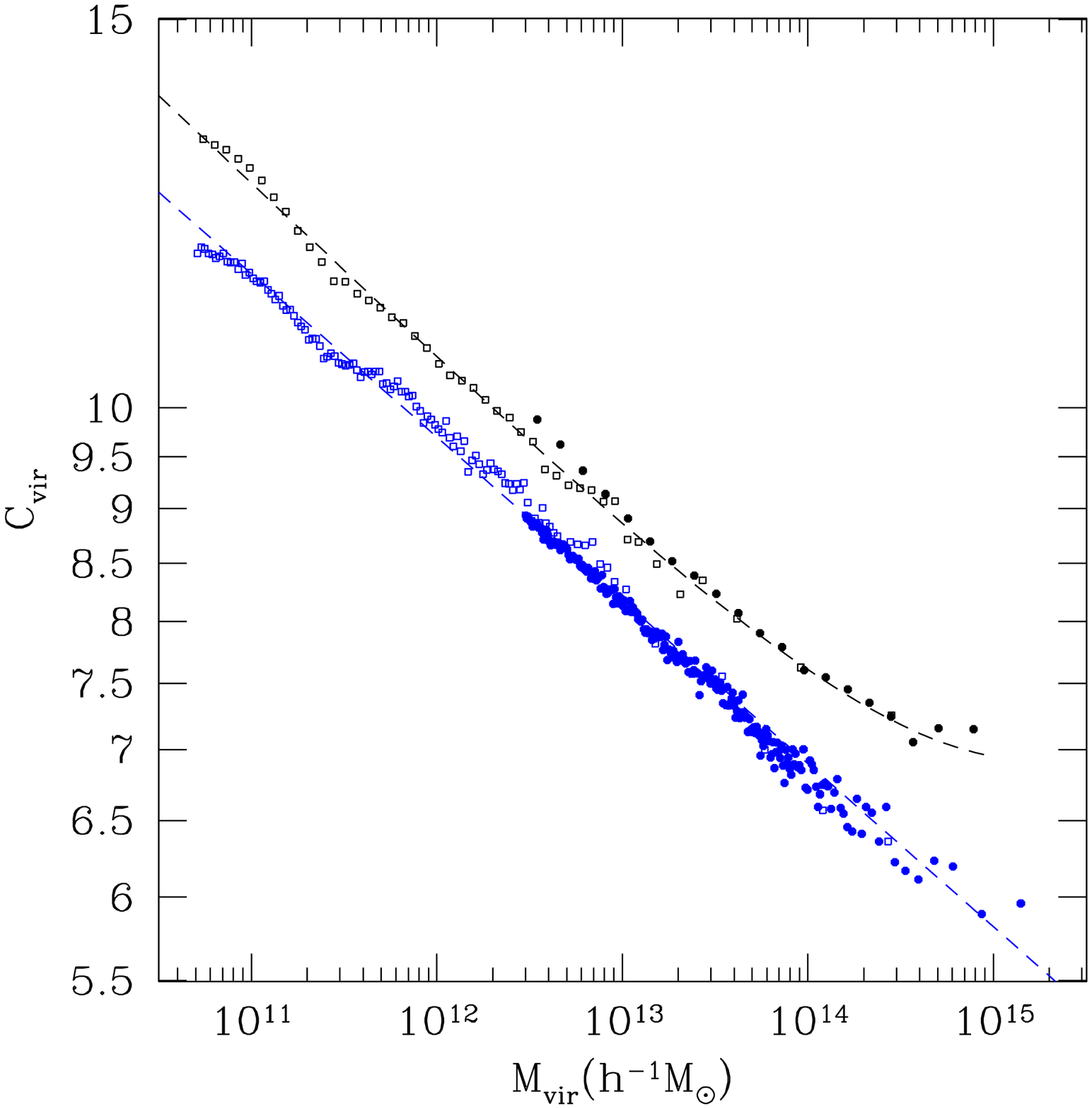}
\caption{Effects of halo selection. The left panel
  shows results for the overdensity 200 values; the right panel is for
  the virial overdensity. Bottom symbols and curves show
  concentrations for all halos selected by mass. The top symbols and
  curves are for relaxed halos selected by maximum circular
  velocity. Open symbols show results from the Bolshoi simulation;
  filled symbols are for the Multidark simulation. All halos selected
  by $\Vmax$ are in between the two sets of curves.}
\label{fig:EffectsA}
\end{figure*}



The selection conditions used in the main part of this paper are: (1)
all halos regardless of their relaxation status; (2) overdensity limit
of 200 relative to the critical density in eq.(\ref{eq:Delta}) defines the
radius of a halo. In the main part of the paper we use notation $c$
for concentrations defined in this way. In the Appendix we also use
another definition of the virial radius. So, in order to distinguish
these two definitions we will use $c_{200}$ for the overdensity 200
definition and $c_{\rm vir}$ for the radius defined by the solution of
the top-hat collapse model in the $\LCDM$ cosmology as approximated
by \citet{BN98}.  For the parameters of the Bolshoi and MultiDark
simulations $\Delta_{\rm vir}= 97$ at $z=0$.
Relative to the matter density this corresponds to $\tilde\Delta
=\Delta/\Omatter = 360$.

One of the interesting features in halo concentration is the upturn in
concentrations at very large masses. Because the largest halos also
grow very fast, one may wonder whether the upturn is just a
non-equilibrium feature. We address this issue by comparing
concentrations of all halos with the concentrations of only relaxed
halos at $z=2$.  Results presented in Figure~\ref{fig:EffectsB} show
that concentrations of relaxed halos are larger than for all halos by
about 10\% and more importantly that the upturn is also clearly present in relaxed
halos.

In Figure~\ref{fig:EffectsA} we investigate effects of selection by
mass, effects of radius definition, and effects of relaxation at $z=0$. On both panels the bottom
curves are for all halos selected by mass and the top curves are for
relaxed halos selected by $\Vmax$. For the latter we show average mass
for halos selected by velocities. Results for all halos selected by
$\Vmax$ are in between the two curves and are not shown to avoid
crowding. In agreement with previous results
\citep[e.g.,][]{Neto07,Maccio08}, we find that concentrations of
relaxed halos are larger than those of all halos. For smaller masses
$M<10^{13}\Msunh$ the two concentrations have nearly the same slopes
and differ by $\sim 7-10\%$. The difference increases at larger masses
where the concentration of relaxed halos practically stops declining
at $M\approx 10^{15}\Msunh$.

The following approximations provide fits for halo concentrations at $z=0$
for masses $M=5\times 10^{10}-10^{15}\Msunh$: \medskip
\noindent Relaxed halos selected by $\Vmax$:
\begin{equation}
 c_{200} = 7.80\left( \frac{\M200}{10^{12}\Msunh} \right)^{-0.08}
               \left[1+ 0.2\left( \frac{\M200}{10^{15}\Msunh} \right)^{1/2}\right]
\label{eq:relax200}
\end{equation}
\begin{equation}
 c_{\rm vir} = 10.5\left( \frac{\Mvir}{10^{12}\Msunh} \right)^{-0.08}
               \left[1+ 0.15\left( \frac{\Mvir}{10^{15}\Msunh} \right)^{1/2}\right]
\label{eq:relaxVir}
\end{equation}

\noindent All halos selected by halo mass:
\begin{equation}
 c_{200} = 7.28\left( \frac{\M200}{10^{12}\Msunh} \right)^{-0.074}
\label{eq:all200}    
\end{equation}
\begin{equation}
 c_{\rm vir} = 9.7\left( \frac{\Mvir}{10^{12}\Msunh} \right)^{-0.074}
\label{eq:allVir}    
\end{equation}

In summary, we find that different selection criteria work very
differently: some are important and some are not: 
\begin{itemize} 
\item Differences in
concentrations due to selection of halos by mass or by $\Vmax$ are
small at small masses: for $\M200 < 10^{13}\Msunh$ halos selected by
$\Vmax$ have concentrations larger only by a factor $1.02-1.04$. The
differences somewhat increase at larger masses: at $\M200 =5\times
10^{14}\Msunh$ the difference is a factor of $1.08$.
\item As expected, the overdensity threshold (in our case either
  $200\rho_{\rm cr}$ or the virial overdensity) has a large impact on halo
  concentration. We find that at $z=0$ the ratio of the concentrations
  is nearly independent on halo mass: $c_{\rm vir}/c_{200}
  =1.35$. This is consistent with previous results
  \citep[e.g.][]{Maccio08}. Note that one may naively expect that
  there should be a mass-dependent correction due the change in the
  overdensity threshold \citep{Hu}. However, this is only
  true for a toy model of an isolated halo with the NFW density
  profile. In reality, the situation is more complex and there are two
  effects: one due to the fact that the same halo is measured at
  different radii and another due to the difference in the sets of halos:
  some of distinct halos defined by small $\R200$ radius cease to be
  distinct and become subhalos for large virial radius.
\item Halos with large masses are effected the most by a particular
  choice of halo selection.  For example,  Figure~\ref{fig:EffectsA} indicates that
  the concentration $c_{200}$ of relaxed halos selected by $\Vmax$ is
  20 percent larger than that of all halos selected by mass for $\M200
  =10^{15}\Msunh$.
\end{itemize}

  \bsp

\label{lastpage}

\end{document}